# Complexity and Philosophy

Francis HEYLIGHEN[1], Paul CILLIERS[2], Carlos GERSHENSON[1]

[1] Evolution, Complexity and Cognition, Vrije Universiteit Brussel,
[2] Philosophy Department, University of Stellenbosch

**ABSTRACT**. The science of complexity is based on a new way of thinking that stands in sharp contrast to the philosophy underlying Newtonian science, which is based on reductionism, determinism, and objective knowledge. This paper reviews the historical development of this new world view, focusing on its philosophical foundations. Determinism was challenged by quantum mechanics and chaos theory. Systems theory replaced reductionism by a scientifically based holism. Cybernetics and postmodern social science showed that knowledge is intrinsically subjective. These developments are being integrated under the header of "complexity science". Its central paradigm is the multi-agent system. Agents are intrinsically subjective and uncertain about their environment and future, but out of their local interactions, a global organization emerges. Although different philosophers, and in particular the postmodernists, have voiced similar ideas, the paradigm of complexity still needs to be fully assimilated by philosophy. This will throw a new light on old philosophical issues such as relativism, ethics and the role of the subject.

## Introduction

Complexity is perhaps the most essential characteristic of our present society. As technological and economic advances make production, transport and communication ever more efficient, we interact with ever more people, organizations, systems and objects. And as this network of interactions grows and spreads around the globe, the different economic, social, technological and ecological systems that we are part of become ever more interdependent. The result is an ever more complex "system of systems" where a change in any component may affect virtually any other component, and that in a mostly unpredictable manner.

The traditional scientific method, which is based on analysis, isolation, and the gathering of *complete* information about a phenomenon, is incapable to deal with such complex interdependencies. The emerging science of complexity (Waldrop, 1992; Cilliers, 1998; Heylighen, 1997) offers the promise of an alternative methodology that would be able tackle such problems. However, such an approach needs solid foundations, that is, a clear understanding and definition of the underlying concepts and principles (Heylighen, 2000).

Such a conceptual framework is still sorely lacking. In practice, applications of complexity science use either very specialized, technical formalisms, such as network clustering algorithms, computer simulations and non-linear differential equations, or rather vaguely defined ideas and metaphors, such as emergence and "the edge of chaos". As such, complexity science is little more than an amalgam of methods, models and metaphors from a variety of disciplines rather than an integrated science. Yet, insofar that complexity science can claim a unified focus, it is to be found precisely in its way of thinking, which is intrinsically different from the one of traditional science (Gershenson & Heylighen, 2005).

A basic function of philosophy is to analyse and criticise the implicit assumptions behind our thinking, whether it is based in science, culture or common sense. As such, philosophy can help us to clarify the principles of thought that characterise complexity science and that distinguish it from its predecessors. Vice versa, complexity theory can help philosophy solve some of its perennial problems, such as the origins of mind, organization or ethics. Traditionally, philosophy is subdivided into metaphysics and *ontology*—which examines the fundamental categories of reality, *logic* and *epistemology*—which investigates how we can know and reason about that reality, *aesthetics* and *ethics*.

Aesthetics and ethics link into the questions of value and meaning, which are usually considered to be outside the scope of science. The present essay will therefore start by focusing on the subjects that are traditionally covered by philosophy of science, i.e. the ontology and epistemology underlying subsequent scientific approaches. We will present these in an approximately historical order, starting with the most "classical" of approaches, Newtonian science, and then moving via the successive criticisms of this approach in systems science and cybernetics, to the emerging synthesis that is complexity science. We will then summarise the impact these notions have had in social science and especially (postmodern) philosophy, thus coming back to ethics and other issues traditionally ignored by (hard) science.

**Newtonian science**

Until the early 20th century, classical mechanics, as first formulated by Newton and further developed by Laplace and others, was seen as the foundation for science as a whole. It was expected that the observations made by other sciences would sooner or later be reduced to the laws of mechanics. Although that never happened, other disciplines, such as biology, psychology or economics, did adopt a general *mechanistic* or *Newtonian* methodology and world view. This influence was so great, that most people with a basic notion of science still implicitly equate "scientific thinking" with "Newtonian thinking". The reason for this pervasive influence is that the mechanistic paradigm is compelling by its simplicity, coherence and apparent completeness. Moreover, it was not only very successful in its scientific applications, but largely in agreement with intuition and common-sense. Later theories of mechanics, such as relativity theory and quantum mechanics, while at least as successful in the realm of applications, lacked this simplicity and intuitive appeal, and are still plagued by paradoxes, confusions and multiple interpretations.

The logic behind Newtonian science is easy to formulate, although its implications are subtle. Its best known principle, which was formulated by the philosopher-scientist Descartes well before Newton, is that of *analysis* or *reductionism*: to understand any complex phenomenon, you need to take it apart, i.e. reduce it to its individual components. If these are still complex, you need to take your analysis one step further, and look at their components.

If you continue this subdivision long enough, you will end up with the smallest possible parts, the *atoms* (in the original meaning of "indivisibles"), or what we would now call "elementary particles". Particles can be seen as separate pieces of the same hard, permanent substance that is called matter. Newtonian ontology therefore is *materialistic*: it assumes that all phenomena, whether physical, biological, mental or social, are ultimately constituted of matter.

The only property that fundamentally distinguishes particles is their position in space (which may include dimensions other than the conventional three). Apparently different substances, systems or phenomena are merely different arrangements in space of fundamentally equivalent pieces of matter. Any change, development or evolution is therefore merely a geometrical rearrangement caused by the movement of the components. This movement is governed by deterministic laws of cause and effect. If you know the initial positions and velocities of the particles constituting a system together with the forces acting on those particles (which are themselves determined by the positions of these and other particles), then you can in principle predict the further evolution of the system with complete certainty and accuracy. The trajectory of the system is not only determined towards the future, but towards the past: given its present state, you can in principle reverse the evolution to reconstruct any earlier state it has gone through.

The elements of the Newtonian ontology are matter, the absolute space and time in which that matter moves, and the forces or natural laws that govern movement. No other fundamental categories of being, such as mind, life, organization or purpose, are acknowledged. They are at most to be seen as epiphenomena, as particular arrangements of particles in space and time.

Newtonian epistemology is based on the reflection-correspondence view of knowledge (Turchin, 1990): our knowledge is merely an (imperfect) reflection of the particular arrangements of matter outside of us. The task of science is to make the mapping or correspondence between the external, material objects and the internal, cognitive elements (concepts or symbols) that represent them as accurate as possible. That can be achieved by simple observation, where information about external phenomena is collected and registered, thus further completing the internal picture that is taking shape. In the limit, this should lead to a perfect, objective representation of the world outside us, which would allow us to accurately predict all phenomena.

All these different assumptions can summarized by the principle of *distinction conservation* (Heylighen, 1990): classical science begins by making as precise as possible distinctions between the different components, properties and states of the system under observation. These distinctions are assumed to be absolute and objective, i.e. the same for all observers. The evolution of the system conserves all these distinctions, as distinct initial states are necessarily mapped onto distinct subsequent states, and vice-versa (this is equivalent to the principle of causality (Heylighen, 1989)). In particular, distinct entities (particles) remain distinct: there is no way for particles to merge, divide, appear or disappear. In other words, in the Newtonian world view there is no place for novelty or creation (Prigogine & Stengers, 1984): everything that exists now has existed from the beginning of time and will continue to exist, albeit in a somewhat different configuration. Knowledge is nothing more than another such distinction-conserving mapping from object to subject: scientific discovery is not a creative process, it is merely an "uncovering" of distinctions that were waiting to be observed.

In essence, the philosophy of Newtonian science is one of *simplicity*: the complexity of the world is only apparent; to deal with it you need to analyse phenomena into their simplest components. Once you have done that, their evolution will turn out to be perfectly regular, reversible and predictable, while the knowledge you gained will merely be a reflection of that pre-existing order.

### Rationality and modernity

Up to this point, Newtonian logic is perfectly consistent—albeit simplistic in retrospect. But if we moreover want to include human agency, we come to a basic contradiction between our intuitive notion of free will and the principle of determinism. The only way Newtonian reasoning can be extended to encompass the idea that people can act purposefully is by postulating the independent category of *mind*. This reasoning led Descartes to propose the philosophy of *dualism*, which assumes that while material objects obey mechanical laws, the mind does not. However, while we can easily conceive the mind as a passive receptacle registering observations in order to develop ever more complete knowledge, we cannot explain how the mind can freely act upon those systems without contradicting the determinism of natural law. This explains why classical science ignores all issues of ethics or values: there simply is no place for purposeful action in the Newtonian world view.

At best, economic science has managed to avoid the problem by postulating the principle of rational choice, which assumes that an agent will always choose the option that maximises its *utility*. Utility is supposed to be an objective measure of the degree of value, "happiness" or "goodness" produced by a state of affairs. Assuming perfect information about the utility of the possible options, the actions of mind then become as determined or predictable as the movements of matter. This allowed social scientists to describe human agency with most of the Newtonian principles intact. Moreover, it led them to a notion of linear progress: the continuous increase in global utility (seen mostly as quantifiable, material welfare) made possible by increases in scientific knowledge. Although such directed change towards the greater good contradicts the Newtonian assumption of reversibility, it maintains the basic assumptions of determinism, materialism and objective knowledge, thus defining what is often called the project of *modernity*.

The assumptions of determinism and of objective, observer-independent knowledge have been challenged soon after classic mechanics reached its apex, by its successor theories within physics: quantum mechanics, relativity theory, and non-linear dynamics (chaos theory). This has produced more than half a century of philosophical debate, resulting in the conclusion that our scientific knowledge of the world is fundamentally *uncertain* (Prigogine & Stengers, 1997). While the notion of uncertainty or indeterminacy is an essential aspect of the newly emerging world view centring around complexity (Gershenson & Heylighen, 2005; Cilliers, 1998), it is in itself not complex, and the physical theories that introduced it are still in essence reductionist. We will therefore leave this aspect aside for the time being, and focus on complexity itself.

**Systems science**

*Holism and emergence*

The first challenges to reductionism and its denial of creative change appeared in the beginning of the twentieth century in the work of process philosophers, such as Bergson, Teilhard, Whitehead, and in particular Smuts (1926), who coined the word *holism* which he defined as the tendency of a whole to be greater than the sum of its parts. This raises the question what precisely it is that the whole has more.

In present terminology, we would say that a whole has *emergent* properties, i.e. properties that cannot be reduced to the properties of the parts. For example, kitchen salt (NaCl) is edible, forms crystals and has a salty taste. These properties are completely different from the properties of its chemical components, sodium (Na) which is a violently reactive, soft metal, and chlorine (Cl), which is a poisonous gas. Similarly, a musical piece has the properties of rhythm, melody and harmony, which are absent in the individual notes that constitute the piece. A car has the property of being able to drive. Its individual components, such as motor, steering wheel, tires or frame, lack this property. On the other hand, the car has a weight, which is merely the sum of the weights of its components. Thus, when checking the list of properties of the car you are considering to buy, you may note that "maximum speed" is an emergent property, while "weight" is not.

In fact, on closer scrutiny practically all of the properties that matter to us in everyday-life, such as beauty, life, status, intelligence..., turn out to be emergent. Therefore, it is surprising that science has ignored emergence and holism for so long. One reason is that the Newtonian approach was so successful compared to its non-scientific predecessors that it seemed that its strategy of reductionism would sooner or later overcome all remaining obstacles. Another reason is that the alternative, holism or emergentism, seemed to lack any serious scientific foundation, referring more to mystical traditions than to mathematical or experimental methods.

### General Systems Theory

This changed with the formulation of systems theory by Ludwig von Bertalanffy (1973). The biologist von Bertalanffy was well-versed in the mathematical models used to describe physical systems, but noted that living systems, unlike their mechanical counterparts studied by Newtonian science, are intrinsically *open*: they have to interact with their environment, absorbing and releasing matter and energy in order to stay alive. One reason Newtonian models were so successful in predicting was because they only considered systems, such as the planetary system, that are essentially closed. Open systems, on the other hand, depend on an environment much larger and more complex than the system itself, so that its effect can never be truly controlled or predicted.

The idea of open system immediately suggests a number of fundamental concepts that help us to give holism a more precise foundation. First, each system has an *environment*, from which it is separated by a *boundary*. This boundary gives the system its own *identity*, separating it from other systems. Matter, energy and information are exchanged across that boundary. Incoming streams determine the system's *input*, outgoing streams its *output*. This provides us with a simple way to connect or *couple* different systems: it suffices that the output of one system be used as input by another system. A group of systems coupled via different input-output relations forms a *network*. If this network functions in a sufficiently coherent manner, we will consider it as a system in its own right, a *supersystem*, that contains the initial systems as its *subsystems*.

From the point of view of the new system, a subsystem or component should be seen not as an independent element, but as a particular type of *relation* mapping input onto output. This transformation or processing can be seen as the function that this subsystem performs within the larger whole. Its internal structure or substance can be considered wholly irrelevant to the way it performs that function. For example, the same information processing function may be performed by neurons in the brain, transistors on a chip, or software modules in a simulation. This is the view of a system as a "black box" whose content we do not know—and do not need to know. This entails an ontology completely different from the Newtonian one: the building blocks of reality are not material particles, but abstract relations, and the complex organizations that together they form. In that sense, systems ontology is reminiscent of the relational philosophy of Leibniz, who had a famous debate with Newton about the assumptions behind the mechanistic world view, but who never managed to develop his philosophical alternative into a workable scientific theory.

By making abstraction of the concrete substance of components, systems theory can establish *isomorphisms* between systems of different types, noting that the network of relations that defines them are the same at some abstract level, even though the systems at first sight belong to completely different domains. For example, a society is in a number of respects similar to a living organism, and a computer to a brain. This allowed von Bertalanffy to call for a *General* Systems Theory, i.e. a way of investigating systems independently of their specific subject domain. Like Newtonian science, systems science strives towards a unification of all the scientific disciplines—from physics to biology, psychology and sociology—but by investigating the patterns of organization that are common to different phenomena rather than their common material components.

Every system contains subsystems, while being contained in one or more supersystems. Thus, it forms part of a *hierarchy* which extends upwards towards ever larger wholes, and downwards towards ever smaller parts (de Rosnay, 1979). For example, a human individual belongs to the supersystem "society" while having different organs and physiological circuits as its subsystems. Systems theory considers both directions, the downward direction of reduction or analysis, and the upward direction of holism or emergence, as equally important for understanding the true nature of the system. It does not deny the utility of the analytical method, but complements it by adding the integrative method, which considers the system in the broader context of its relations with other systems together with which it forms a supersystem.

Also the concept of emergent property receives a more solid definition via the ideas of *constraint* and *downward causation*. Systems that through their coupling form a supersystem are constrained: they can no longer act as if they are independent from the others; the supersystem imposes a certain coherence or coordination on its components. This means that not only is the behavior of the whole determined by the properties of its parts ("upwards causation"), but the behavior of the parts is to some degree constrained by the properties of the whole ("downward causation" (Campbell, 1974)). For example, the behavior of an individual is controlled not only by the neurophysiology of her brain, but by the rules of the society to which she belongs.

Because of the dependencies between components, the properties of these components can no longer vary independently: they have to obey certain relationships. This makes much of the individual properties irrelevant, while shifting the focus to the state of their relationship, which will now define a new type of "emergent" property. For example, a sodium atom that gets bonded to a chlorine atom, forming a salt molecule, loses its ability to react with other atoms, such as oxygen, but acquires the ability to align itself into a crystalline structure with other salt molecules.

*Cybernetics and the subjectivity of knowledge*

Tight relationships between subsystems turn the whole into a coherent organization with its own identity and autonomy. Cybernetics, an approach closely associated to systems theory, has shown how this autonomy can be maintained through goal-directed, apparently intelligent action (Ashby, 1964; Heylighen & Joslyn, 2001). The principle is simple: certain types of circular coupling between systems can give rise to a negative feedback loop, which suppresses deviations from an equilibrium state. This means that the system will actively compensate perturbations originating in its environment in order to maintain or reach its "preferred" state of affairs. The greater the variety of perturbations the system has to cope with, the greater the variety of compensating actions it should be able to perform (Ashby's (1964) law of requisite variety), and the greater the knowledge or intelligence the system will need in order to know which action to perform in which circumstances. Research in cybernetics—and later in neural networks, artificial intelligence and cognitive science—has shown how such intelligence can be realized through an adaptive network of relations transforming sensory input into decisions about actions (output). Thus, the systems perspective has done away with the Cartesian split between mind and matter: both are merely particular types of relations.

However, this perspective entails a new view on epistemology. According to cybernetics, knowledge is intrinsically subjective; it is merely an imperfect tool used by an intelligent agent to help it achieve its personal goals (Heylighen & Joslyn, 2001; Maturana & Varela, 1992). Such an agent not only does not need an objective reflection of reality, it can never achieve one. Indeed, the agent does not have access to any "external reality": it can merely sense its inputs, note its outputs (actions) and from the correlations between them induce certain rules or regularities that seem to hold within its environment. Different agents, experiencing different inputs and outputs, will in general induce different correlations, and therefore develop a different knowledge of the environment in which they live. There is no objective way to determine whose view is right and whose is wrong, since the agents effectively live in different environments ("Umwelts")—although they may find that some of the regularities they infer appear to be similar.

This insight led to a new movement within the cybernetics and systems tradition that calls itself "second-order cybernetics" (von Foerster, 1979; Heylighen & Joslyn, 2001). Its main thesis is that we, as observers, are also cybernetic systems. This means that our knowledge is a subjective construction, not an objective reflection of reality. Therefore, the emphasis has to shift from the apparently objective systems around us to the cognitive and social processes by which we construct our subjective models of those systems. This constitutes a major break with traditional systems theory, which implicitly assumed that there is an objective structure or organization in the systems we investigate (Bunge, 1979). This departure was reinforced by the concepts of autonomy, autopoiesis (Maturana & Varela, 1979) and self-organization, that were introduced to characterise natural, living systems in contrast to artificial, engineered systems. These imply that the structure of a system is not given, but developed by the system itself, as a means to survive and adapt to a complex and changing environment.

The rift became even larger when it became clear that many systems, and in particular social systems, do not have any clear structure, function or organization, but consist of a tangle of partly competing, partly co-operating, or simply mutually ignoring subsystems. For example, whereas the older generation of systems thinkers (e.g. Parsons, 1991) viewed society as a stable, organism-like system, where the different subsystems have clearly defined functions in contributing to the common good, the newer generation of social scientists saw an anarchy of conflicting forces with different coalitions and subcultures emerging and disappearing again. In such systems, there are many relationships which cut across apparently hierarchical layers so that a system that is subordinate to another system in one respect, appears superordinate in another respect, an ill-defined configuration that is sometimes called "heterarchy".

The growing awareness of these two limitations to the systems view—the subjectivity of knowledge and the lack of order in autonomous and especially social systems—promoted the emergence of a new science of complex systems in parallel with a "Postmodern" philosophy (Cilliers, 1998).

**Complexity Science**

In the 1980's, a new approach emerged which is usually labelled as *complex adaptive systems* (Holland, 1996) or, more generally, *complexity science* (Waldrop, 1992). Although its origins are largely independent from systems science and cybernetics, complexity science offers the promise to extend and integrate their ideas, and thus develop a radical, yet workable alternative to the Newtonian paradigm. The roots of the complexity movement are diverse, including:

- non-linear dynamics and statistical mechanics—two offshoots from Newtonian mechanics—which noted that the modelling of more complex systems required new mathematical tools that can deal with randomness and chaos;
- computer science, which allowed the simulation of systems too large or too complex to model mathematically;
- biological evolution, which explains the appearances of complex forms through the intrinsically unpredictable mechanism of blind variation and natural selection;
- the application of these methods to describe social systems in the broad sense, such as stock markets, the Internet or insect societies, where there is no predefined order, although there are emergent structures.

Given these scientific backgrounds, most complexity researchers have not yet reflected about the philosophical foundations of their approach—unlike the systems and cybernetics researchers. As such, many still implicitly cling to the Newtonian paradigm, hoping to discover mathematically formulated "laws of complexity" that would restore some form of absolute order or determinism to the very uncertain world they are trying to understand. However, we believe that once the insights from systems science and postmodern philosophy will have been fully digested, a philosophy of complexity will emerge that is truly novel, and whose outline we can at present only vaguely discern.

What distinguishes complexity science is its focus on phenomena that are characterized neither by order—like those studied in Newtonian mechanics and systems science, nor by disorder—like those investigated by statistical mechanics and Postmodern social science, but that are situated somewhere in between, in the zone that is commonly (though perhaps misleadingly) called the *edge of chaos* (Langton, 1990). Ordered systems, such as a crystal, are characterized by the fact that their components obey strict rules or constraints that specify how each component depends on the others. Disordered systems, such as a gas, consist of components that are independent, acting without any constraint. Order is simple to model, since we can predict everything once we know the initial conditions and the constraints. Disorder too is simple in a sense: while we cannot predict the behavior of individual components, statistical independence means that we can accurately predict their *average* behavior, which for large numbers of components is practically equal to their overall behavior. In a truly complex system, on the other hand, components are to some degree independent, and thus autonomous in their behavior, while undergoing various direct and indirect interactions. This makes the global behavior of the system very difficult to predict, although it is not random.

### *Multi-agent systems*

This brings us to the most important conceptual tool introduced by complexity science: the *complex adaptive system,* as defined by Holland (1996), which is presently more commonly denoted as a *multi-agent system*. The basic components of a complex adaptive system are called *agents*. They are typically conceived as "black box" systems, meaning that we know the rules that govern their individual behavior, but we do not care about their internal structure. The rules they follow can be very simple or relatively complex; they can be deterministic or probabilistic. Intuitively, agents can be conceived as autonomous individuals who try to achieve some personal goal or value ("utility" or "fitness") by acting upon their environment—which includes other agents. But an agent does not need to exhibit intelligence or any specifically "mental" quality, since agents can represent systems as diverse as people, ants, cells or molecules. In that respect, complexity science has assimilated the lessons from cybernetics, refusing to draw any a priori boundary between mind and matter.

From evolutionary theory, complexity science has learned that agents typically are ignorant about their wider environment or the long-term effects of their actions: they reach their goals basically by trial-and-error, which is equivalent to *blind variation* followed by the *natural selection* of the agents, actions or rules for action that best achieve fitness. Another way to describe this short-sightedness is by noting that agents are intrinsically egocentric or *selfish*: they only care about their own goal or fitness, initially ignoring other agents. Only at a later stage may they "get to know" their neighbours well enough to develop some form of cooperation (e.g. Axelrod, 1984). But even when the agents are intelligent and knowledgeable enough to select apparently rational or cooperative actions, they—like us—are intrinsically *uncertain* about the remote effects of their actions.

This limited range of rational anticipation is reflected at the deepest level by the principle of *locality*: agents only interact with (and thus get the chance to "know") a small number of other agents which form their local neighbourhood. Yet, in the longer term these local actions typically have global consequences, affecting the complex system as a whole. Such global effects are by definition unexpected at the agent level, and in that sense *emergent*: they could not have been inferred from the local rules (properties) that determine the agents' behavior. For us as outside observers, such emergent properties do not necessarily come as a surprise: if the interactions between the agents are sufficiently regular or homogeneous, as in the interactions between molecules in a crystal or a gas, we may be able to predict the resulting global configuration. But in the more general cases, it is impossible to extrapolate from the local to the global level.

This may be better understood through the following observations. First, agents' goals are intrinsically independent, and therefore often in conflict: the action that seems to most directly lead to A's goal, may hinder B in achieving its goal, and will therefore be actively resisted by B. This is most obvious in economies and ecosystems, where individuals and organisms are always to some degree competing for resources. Eating a zebra may be an obvious solution to the lion's problem of hunger, but that action will be resisted by the zebra. Increasing the price may be the most obvious way for a producer to increase profit, but that will be resisted by the clients switching to other suppliers. Such inherent conflicts imply that there is no "global optimum" for the system to settle in, i.e. an equilibrium state that maximally satisfies *all* agents' goals. Instead, agents will *co-evolve*: they constantly adapt to the changes made by other agents, but through this modify the others' environment, thus forcing them to adapt as well (cf. Kauffman, 1995). This results in an on-going process of mutual adaptation, which in biology is elegantly expressed by metaphors such as an "arms race" or the "Red Queen principle".

Second, since actions are local, their effects can only propagate step by step to more remote agents, thus diffusing across the whole network formed by the agents and their relationships of interaction. The same action will in general have multiple effects in different parts of the network at different times. Some of those causal chains will close in on themselves, feeding back into the conditions that started the chain. This makes the system intrinsically *non-linear*. This means that there is no proportionality between cause and effect. On the one hand, small fluctuations may be amplified to large, global effects by positive feedback or "autocatalysis". Such sensitive dependence on initial conditions, which is often referred to as the "butterfly effect", is one of the hallmarks of deterministic *chaos*, i.e. globally unpredictable changes produced by locally deterministic processes. But complex systems don't need to be deterministic to behave chaotically. On the other hand, feedback can also be negative, so that large perturbations are suppressed, possibly resulting in the stabilisation of a global configuration.

*Creative evolution*

The combination of these different effects leads to a global evolution that is not only unpredictable, but truly creative, producing emergent organization and innovative solutions to global and local problems. When we focus on the complex system in itself, we can call the process *self-organization*: the system spontaneously arranges its components and their interactions into a sustainable, global structure that tries to maximize overall fitness, without need for an external or internal designer or controller (Heylighen, 2002; Kauffman, 1995). When we focus on the relation between the system and the environment, we may call it *adaptation* (Holland, 1996): whatever the pressures imposed by the environment, the system will adjust its structure in order to cope with them. Of course, there is no guarantee of success: given the intrinsic sensitivity and unpredictability of the system, failures and catastrophes can (and do) happen, often when we do not expect them. But in the long term, on-going self-organization and adaptation appear to be the rule rather than the exception.

As such, the complexity paradigm answers a fundamental philosophical question that was left open by earlier approaches: what is the origin of the order, organization and apparent intelligence that we see around us (Heylighen, 2000)? Newtonian and systems science had eluded that question by considering that order as pre-existing. Earlier, pre-scientific philosophies had tackled the question by postulating a supernatural Creator. Darwin's theory of evolution through natural selection had provided a partial answer, which moreover remained restricted to biological systems, and thus is considered unsatisfactory by many. The co-evolution of many, interacting agents, on the other hand, seems able to explain the emergence of organization in any domain or context: physical, chemical, biological, psychological or social.

While it is difficult to imagine the limitless ramifications of such a process without the support of complex computer simulations or mathematical models, the basic principle is simple: each agent through trial-and-error tries to achieve a situation that maximises its fitness within the environment. However, because the agent cannot foresee all the consequences, actions will generally collide with the actions of other agents, thus reaping a less than optimal result. This pressures the agent to try out different action patterns, until one is found that reduces the friction with neighbouring agents' activities, and increases their synergy. This creates a small, relatively stable "community" of mutually adapted agents within the larger collective. Neighbouring agents too will try to adapt to the regime of activity within the community so that the community grows. The larger it becomes, the stronger its influence or "selective pressure" on the remaining agents, so that eventually the whole collective will be assimilated into the new, organized regime. Whenever the organization encounters a problem (loss of fitness), whether because of internal tensions or because of perturbations from the outside, a new adaptation process will be triggered in the place where the problem is experienced, propagating as far as necessary to absorb all the negative effects.

In such an organized collective, individual agents or agent communities will typically specialise in a particular activity (e.g. processing a particular type of resource) that complements the activities of the other agents. As such, agents or communities can be seen to fulfil a certain function or role within the global system, acting like functional subsystems. Thus, complex adaptive systems may come to resemble the supersystems studied by systems theory. Such a supersystem can be seen as an agent at a higher level, and the interaction of several such "superagents" may recursively produce systems at an ever higher hierarchical level (Heylighen, 2002).

However, the organization of such a complex system is not frozen, but flexible, and the same agent may now seem to participate in one function, then in another. In some cases, like in multicellular organisms, the functional differentiation appears pretty stable. In others, like in our present society or in the brain, agents regularly switch roles. But the difference is merely one of degree, as all complex systems created through self-organization and evolution are intrinsically adaptive, since they cannot rely on a fixed plan or blueprint to tell them how they should behave. This makes a naturally evolved organization, such as the brain, much more *robust* than an organization that has been consciously designed, such as a computer. The intrinsic uncertainty, which appeared like a weakness, actually turns out to be a strength, since it forces the system to have sufficient reserves or redundancy and to constantly try out new things so as to be prepared for any eventuality.

**Complexity and (Postmodern) Philosophy**

Although ideas from complexity theory have had a substantial impact on various disciplines outside the "hard" sciences from where they originated, in particular in sociology (e.g. Urry, 2003, Byrne, 1998) and organisational sciences (e.g. Stacey *et al* 2000, Stacey, 2001, Richardson, 2005), the impact on mainstream philosophy has not been as significant as one would expect. This is surprising given that the related domains of cognitive science and evolutionary theory have inspired plenty of philosophical investigations.

One reason may be that the Anglo-Saxon tradition of "analytic" philosophy by its very focus on analysing problems into their logical components is inimical to the holism, uncertainty and subjectivity entailed by complexity. Within the English-speaking academic world, we only know two philosophers who have founded their ontology on the holistic notion of system: Bunge (1979), who otherwise remains a believer in objective, logical knowledge, and Bahm (1987), who continues the more mystical tradition of process philosophy. The few philosophers, such as Morin (1992), Luhmann (1995) and Stengers (Prigogine & Stengers, 1984; 1997), who have directly addressed complexity, including the uncertainty and subjectivity that it entails, all seem to come from the continental tradition.

Another reason may be that much of complexity theory has resulted from developments in mathematics and computational theory. This is not the normal domain of most philosophers. Complexity has therefore been mostly discussed in philosophy of science, mathematics and computation, but not really in philosophy of culture and social philosophy. To the extent that it has, the discussion either ignored a lot of already established work on complexity (e.g. Rescher, 1998), or made use of ideas derived mainly from chaos theory, something we regard as a very limited subset of complexity studies in general (e.g. Taylor, 2003). (A number of insightful and stimulating papers, focusing to a large extent on the work of Luhmann, can be found in *Observing Complexity*, edited by Rasch and Wolfe (2000). The paper by Rasch himself, entitled *Immanent Systems, Transcendental Temptations, and the Limits of Ethics* is of particular interest.)

A further reason may be that philosophy has somehow always been engaged with complex issues, even if it has not been done in the language used by contemporary complexity theorists. If this is true, the language of complexity could fruitfully inform a number of philosophical debates and, *vice versa*, ideas from philosophy of language, culture and society could enrich discussions on complexity as such. To an extent this interaction is taking place in that part of philosophy sometimes characterised as "postmodern". (Note that this term should be used with caution. It can refer to a very wide range of positions, sometimes pejoratively and sometimes merely as a verbalism. It will not be used here to refer to flabby or relativist positions, but to a number of solid philosophical positions critical of foundational forms of modernism.)

The general sensitivity to complexity in philosophy can be traced in an interesting way by looking at positions incorporating a systems perspective. A good starting point would be Hegel. The dialectical process whereby knowledge, and the relationship between knowledge and the world, develops, works, for Hegel, in a systemic way. A new synthesis incorporates the differences of the thesis and the antithesis, but it already poses as a new thesis to be confronted. Thus Hegel's system is an historical entity, something with a procedural nature. The problem is that, for Hegel, this is a converging process, ultimately culminating in what Cornell (1992) calls a "totalising system". His position thus remains fully within the modernist paradigm.

Several philosophical positions incorporate important insights from Hegel, but resist this idea of convergence. A good example is Adorno's negative dialectics, where the dialectical process drives a *diverging* process (see Held, 1980). More influential examples, in terms of the complexity debate at least, are the systems theories of Freud (1950) and Saussure (1974). In his early *Project for a Scientific Psychology*, Freud develops a model of the brain based on a system of differences which is structurally equivalent to Saussure's model of language. In this understanding, signs in a system do not have meaning on their own, but through the relationships amongst all the signs in the system. The work of Freud and Saussure, especially in the way it has been transformed and elaborated by thinkers like Derrida and Lacan, has been central to much of postmodern philosophy, as discussed by Cilliers (1998: 37-47).

Modernism can be characterised, in Lyotard's words (1988: xxiv), as a search for a single coherent meta-narrative, i.e. to find the language of the world, the one way in which to describe it correctly and completely. This can only be a reductive strategy, something which reduces the complexity and the diversity of the world to a finite number of essential features. If the central argument of postmodernism is a rejection of this dream of modernism, then postmodernism can be characterised in general as a way of thinking which is sensitive to the complexity of the world. Although he does not make use of complexity theory as such, Derrida is sensitive to exactly this argument. "If things were simple, word would have gotten around" he famously says in the Afterword to *Limited Inc* (Derrida, 1988: 119). Lyotard's (1988) characterisation of different forms of knowledge, and his insistence on what he calls "paralogy", as opposed to conventional logic, is similarly an acknowledgement of the complexity of the postmodern world (see Cilliers, 1998, p.112-140, for a detailed discussion of Lyotard's position from within a complexity perspective). An innate sensitivity to complexity is also central to the work of Deleuze and Guattari (1987; Guattari, 1995). Many of their post-Freudian insights, and especially the idea of the "rhizome" deny reductive strategies. Their work has also been interpreted specifically from a complexity perspective (DeLanda, 2005; Ansell-Pearson, 1999).

As yet, applications of complexity theory to the social sciences have not been very productive. There may be a number of reasons for this, but it can be argued that many social theorists were introduced to complexity via the work done by "hard" complexity scientists, perhaps mostly through the work of what one can broadly call the Santa Fe school (Waldrop, 1992). Since this work is strongly informed by chaos theory, it contains strong reductive elements, and in that sense it is still very much "modernist" in flavour. The "postmodern" approach, especially one informed by recent developments in general complexity theory, could be extremely useful in enriching the discourse on social and cultural complexity. There are, without any doubt, a number of postmodern positions which are just too flaky to take seriously, but the all too common knee-jerk rejection of anything labelled "postmodern" —irrespective of whether this label is used correctly or not—will have to be tempered in order to get this discourse going. Space does not allow a detailed discussion of the different themes which could form part of this discourse, but a few can be mentioned briefly.

### *The Structure of Complex Systems*

The emphasis on ideas from chaos theory has negatively influenced our understanding of the structure of complex systems. Most natural complex systems have a well-defined structure and they are usually quite robust. Despite their non-linear nature, they are not perpetually balanced on a knife's edge. Theories of meaning derived from a post-structural understanding of language, e.g. deconstruction, could illuminate this debate. In order for this illumination to take place, it will first have to be acknowledged that deconstruction does not imply that meaning is relative. (See Cilliers, 2005.)

### *Boundaries and Limits*

The relationship between a complex system and its environment or context is in itself a complex problem. When dealing with social systems, it is often unclear where the boundary of a system is. It is often a matter of theoretical choice. Furthermore, the notion "limit" is often confused with the notion "boundary", especially where the theory of autopoiesis is used. The problems of "framing" and the way in which context and system mutually constitute each other could be elaborated on from several postmodern viewpoints. (See Cilliers, 2001.)

*The Problem of Difference*

For the modernist, difference and diversity was always a problem to be solved. For the postmodernist, diversity is not a problem, but the most important resource of a complex system. Important discussions on diversity and difference, including issues in multi-culturalism, globalisation, bio-diversity, sustainability and the nature of social systems in general, could benefit greatly from the work done on difference by Saussure, Derrida, Deleuze and other post-structural thinkers.

*The Idea of the Subject*

The Enlightenment idea of a self-contained, atomistic subject is undermined in similar ways by complexity theories and postmodernism. Nevertheless, the idea of the subject cannot be dismissed. Notions of agency and responsibility remain extremely important, but they have to be supplemented with insights from theories of self-organisation and social construction. There are many unresolved problems in this area and some very exciting work could be done here. For a very preliminary attempt, see Cilliers and De Villiers (2000).

*Complexity and Ethics*

Moral philosophy has been strongly influenced by the modernist ideal of getting it exactly right. Complexity theory argues that, since we cannot give a complete description of a complex system, we also cannot devise an unchanging and non-provisional set of rules to control the behaviour of that system. Complexity theory (and postmodernism), of course, cannot devise a better ethical system, or at least not a system that will solve the problem. What it can do however, is to show that when we deal with complexity—and in the social and human domain we always do—we cannot escape the moment of choice, and thus we are *never* free of normative considerations. Whatever we do has ethical implications, yet we cannot call on external principles to resolve our dilemmas in a final way. The fact that some form of ethics is unavoidable seems to be a very important insight from complexity theory. This follows as well from evolutionary-cybernetic reasoning (Heylighen, 2000; Turchin, 1990) as from the classic multi-agent simulations of the emergence of cooperation (Axelrod, 1984). This resonates strongly with post-structural and Derridean ethics (see Cilliers, 1998: 136–140, Cilliers 2004)

*Complexity and Relativism*

If complexity theory ultimately argues for the incompleteness of knowledge, it becomes a target, just like postmodernism, for those accusing it of relativism. This is not a meaningful accusation and has led to a lot of fruitless debates (cf. Sokal's hoax). The dismissal of positions which try to be conscious of their own limitations is often a macho, if not arrogant move, one which is exactly insensitive to the ethical dimension involved when we deal with complexity. Modest positions do not have to be weak ones (see Norris, 1997; Cilliers, 2005). The development of a theoretical position which moves beyond the dichotomy of relativism and foundationalism (two sides of the same coin) is vital (cf. Heylighen, 2000).

The intersection between complexity and postmodern philosophy could lead to both exciting and very useful research. One of the rewards of this approach is that it allows insights from both the natural and the social sciences without one having to trump the other.

### Some Current Trends

The contributions to the session on Philosophy and Complexity at the Complexity, Science, and Society conference (University of Liverpool, 2005), that was organized by one of us (Gershenson), and that we all participated in, provided a sample of current trends in the field. It was clear that concepts from complexity have not gone very deeply into philosophy, but the process is underway, since there are many open questions posed by scientific advances related to complexity, affecting especially epistemology and ethics. For example, research in life sciences demands a revaluation of our concept of 'life', while studies in cognitive sciences question our models of 'mind' and 'consciousness'.

The terminology introduced by complexity has already propagated, but not always with the best results. For example, the concept of emergence is still not well understood, a situation fuelled by the ignorant abuse of the term, although it is slowly being demystified.

An important aspect of complex adaptive systems that is currently influencing philosophy is that of evolution. The dynamism introduced by cybernetics and postmodernism has not yet invaded all its possible niches, where remnants of reductionism or dualism remain. Philosophy no longer is satisfied by explaining why something is the way it is, but it needs to address the question of how it got to be that way.

## Conclusion

For centuries, the world view underlying science has been Newtonian. The corresponding philosophy has been variously called reductionism, mechanicism or modernism. Ontologically, it reduces all phenomena to movements of independent, material particles governed by deterministic laws. Epistemologically, it holds the promise of complete, objective and certain knowledge of past and future. However, it ignores or even denies any idea of value, ethics, or creative processes, describing the universe as merely a complicated clockwork mechanism.

Over the past century, various scientific developments have challenged this simplistic picture, gradually replacing it by one that is complex at the core. First, Heisenberg's uncertainty principle in quantum mechanics, followed by the notion of chaos in non-linear dynamics, showed that the world is intrinsically unpredictable. Then, systems theory gave a scientific foundation to the ideas of holism and emergence. Cybernetics, in parallel with postmodern social science, showed that knowledge is intrinsically subjective. Together with the theories of self-organization and biological evolution, they moreover made us aware that regularity or organization is not given, but emerges dynamically out of a tangle of conflicting forces and random fluctuations, a process aptly summarized as "order out of chaos" (Prigogine & Stengers, 1984).

These different approaches are now starting to become integrated under the heading of "complexity science". Its central paradigm is the multi-agent system: a collection of autonomous components whose local interactions give rise to a global order. Agents are intrinsically subjective and uncertain about the consequences of their actions, yet they generally manage to self-organize into an emergent, adaptive system. Thus, uncertainty and subjectivity should no longer be viewed negatively, as the loss of the absolute order of mechanicism, but positively, as factors of creativity, adaptation and evolution.

Although a number of (mostly postmodern) philosophers have expressed similar sentiments, the complexity paradigm still needs to be assimilated by academic philosophy. This may not only help philosophy solve some of its perennial problems, but help complexity scientists become more aware of the foundations and implications of their models.